\documentclass[12pt]{article}
\usepackage{graphicx}
\usepackage{a4}
\usepackage{amsmath}
\usepackage{amssymb}

\def\R{\mathbb R}				\def\N{\mathbb N}
\def\Z{\mathbb Z}		\def\Q{\mathbb Q}

\def\ie{i.\,e. }				

\def\beq#1\eeq{\begin{equation}#1\end{equation}}
\def\bal#1\eal{\begin{align}#1\end{align}}
\def\bals#1\eals{\begin{subequations}\begin{align}#1%
	\end{align}\end{subequations}}
\def\bse#1\ese{\begin{subequations}#1\end{subequations}}

\newcommand{\torus}{T_\theta^2}
\newcommand{\op}[1]{\hat #1}
\newcommand{\pd}{\partial}

\DeclareMathOperator{\sgn}{sgn}

\begin{document}
\hfill\textsf{UWThPh--2002--30}

\vspace{10mm}
\begin{center}
	\renewcommand{\thefootnote}{\fnsymbol{footnote}}
	\Large The Landau-Problem on the $\theta$-Deformed Two-Torus\vskip5mm
	\large H. Grosse\footnote{\ttfamily harald.grosse@mailbox.univie.ac.at}\;
	and M. Kornexl\footnote{\ttfamily mkornexl@thp.univie.ac.at}
	\footnote{research supported by DOC grant of the Austrian Academy of
	Science}\\
	\it Institute for Theoretical Physics, University of Vienna\\
	\it Boltzmanngasse 5, 1090 Wien, Austria\vskip5mm
	\today
\end{center}\vskip10mm

\begin{abstract}
	We study the Landau problem on the $\theta$-deformed two-torus and use
	well-known projective modules to obtain perturbed energy spectra. For a
	strong magnetic field $B$ the problem can be restricted to one particular
	Landau-level. First we represent generators of the algebra of the
	non-commutative torus $\torus$ as finite dimensional matrices. A second
	approach leads to a reducible representation with a $\theta$-dependent
	center. For a simple periodic potential, the rational part of the
	Hofstadter-butterfly spectrum is obtained.
\end{abstract}

\section{Introduction}
The problem of a charged particle moving in two dimensions with a magnetic
field $B$ applied perpendicular to the two-dimensional manifold is known as the
(two-dimensional) Landau problem. For $B$ very large a perturbation $V(x_i)$ of
the free Hamiltonian $\mathcal H_0 = \frac{1}{2 m} (\pi_1^2 + \pi_2^2)$ can be
treated by projecting onto one Landau level. Due to this perturbation, the
degeneracy of the levels will be lifted, but, since the energy gap between two
separate Landau levels $\sim B/m$, is very large, the perturbation does not mix
different Landau levels and the projection onto one level is justified.

The procedure of projecting onto one Landau level is known since a long time in
solid state physics as the Peierls substitution \cite{Peierls:1933}.

We look how a $\theta$-deformation of the underlying manifold effects the
Peierls substitution. We use the algebra over the $\theta$-deformed plane,
$\mathcal A_\theta$, generated by two elements $\op x_1$ and $\op x_2$, which
satisfy $[\op x_1, \op x_2] = 2 \pi i \theta$. The phase-space of
$\theta$-deformed quantum mechanics is generated by the coordinates $\op x_i$
and momenta $\op p_i, i= 1, 2$, subject to the relations
\beq
	[\op x_1, \op x_2] = 2 \pi i \theta, \qquad
	[\op x_k, \op p_l] = i \delta_{k,l} \qquad\text{and}\qquad
	[\op p_1, \op p_2] = 0.
\eeq
Like in the commutative case, a magnetic field is introduced by replacing the
canonical momenta $\op p_i$ by kinetic ones $\op \pi_i$, which are covariant
under local $U(1)$ gauge transformations $g= g(\op x_i)= e^{i \varphi(\op
x_i)}$, \ie $\op p_i \to \op \pi_i = \op p_i - A_i$, with $A_i = A_i (\op x_1,
\op x_2)$ a $U(1)$ gauge potential transforming under $g$ according to $A_j \to
A_j' = g A_j g^{-1} + i g (\pd_j g^{-1})$. Since $\op x_1$ and $\op x_2$ do not
commute, also the coordinates have to be replaced by their covariant
counterparts $\op x_i \to \op \xi_i = \op x_i + 2 \pi \theta \epsilon_{i j}
A_j$. The commutation relations for the covariant phase-space coordinates read
\beq
	[\op \xi_1, \op \xi_2] = 2 \pi i \, \theta (1 + 2 \pi \theta F), \quad
	[\op \xi_k, \op \pi_l] = i (1 + 2 \pi \theta F) \delta_{k, l}
	\quad\text{and}\quad
	[\op \pi_1, \op \pi_2] = i F,
\eeq
with $F= \pd_{[1} A_{2]} - i [A_1, A_2]$, the gauge field strength. For $1 +
2\pi \theta F \ne 0$, these relations are, up to some rescaling, equivalent to
\beq\label{eq:phase-space-CR}
	[x_1, x_2] = 2 \pi i \theta, \qquad
	[x_k, \pi_l] = i \delta_{k,l} \qquad\text{and}\qquad
	[\pi_1, \pi_2] = i B,
\eeq
with $B= F/(1 + 2 \pi \theta F)$.

In chapter \ref{sec:rep-Enm} and \ref{sec:proj-Enm} we use the well-known
projective modules over $\torus$ and take a connection with constant curvature
to obtain representations of the deformed tori algebra. Chapter
\ref{sec:rep-R2} deals with a different reducible representation motivated by
physics. We require that the magnetic translation operators commute with
covariant coupled momenta and obtain a quantization condition. The flux turns
out to be rational. For irrational deformation parameter $\theta$, $B$ of Eq.
\eqref{eq:phase-space-CR} then is always irrational. Translation invariance is
broken for nonzero magnetic flux. Using the results of chapters
\ref{sec:rep-Enm} and \ref{sec:rep-R2}, we calculate the projection of the
generators $U_j = e^{i x_j}, j= 1, 2$ of $\torus$ onto one particular Landau
level. The projected operators $U_j^{(\mu)}$ are represented as finite
dimensional matrices.

In chapter \ref{sec:V_mu} we calculate the energy corrections due to a small
periodic potential $V(x_i) = v (\cos x_1 + \cos x_2)$. This yields the
so-called Hofstadter butterfly \cite{Hofstadter:1976} (or to be more precisely
the rational part, \ie the part arising from rational fluxes, of the Hofstadter
butterfly).

\section{Representation of $\torus$}\label{sec:rep-Enm}
The algebra generated by the elements $U_1$ and $U_2$ subject to the relation
\beq\label{eq:torus-CR}
	U_1 U_2 = e^{-2 \pi i \theta} U_2 U_1
\eeq
is known as the algebra of functions over the non-commutative two-torus $\torus$
with deformation parameter $\theta$. Two unitary operators on a Hilbert space
obeying Eq. \eqref{eq:torus-CR} specify a representation of $\torus$, or, in
other words, a module over $\torus$. Without loss of generality we assume the
periods of the torus to be equal to $2 \pi$. Further we restrict ourselves to
irrational $\theta$ for the moment.

As shown by Connes and Rieffel \cite{Connes:1987,Rieffel:1988}, for $\theta$
irrational, any $\torus$-module is either free (for constant magnetic field
$B$, this corresponds to $B=0$) or isomorphic to the module $E_{n,m}$ for some
integers $n$ and $m$. The elements of the $\torus$-module $E_{n,m}$ will be
Schwartz class functions $\phi_j (x) \in \mathcal S (\R \times \Z_m), x \in \R,
j \in \Z_m$, with $\Z_m$ the cyclic group of order $m$. The action of the
generators $U_i$ of $\phi_j(x)$ is given by
\bals
	(U_1 \phi)_j (x) & = \phi_{j-1} (x - \frac n m - \theta), \\
	(U_2 \phi)_j (x) & = e^{2 \pi i (x - j \, n/m)} \phi_j (x),
\eals
with some integer $n$.

The projective modules are classified by their $K$ theory group which is the
rank two abelian group $\Z^2$. The set of classes of actual finite projective
modules is the cone of positive elements, which for $\Z^2$ is 
\beq
	\{ (x, y) \in \Z^2; x + \theta y > 0 \}.
\eeq
For $E_{n,m}$ the coordinates $x$ and $y$ are
\beq
	x = \sigma n, \qquad y = \sigma m,
\eeq
with $\sigma = \sgn (n + m\theta)$ (c.f. \cite{Connes:1998cr}).

A magnetic field $B$ perpendicular to the torus is introduced via minimal
coupling, \ie the canonical momenta $p_j = -i \pd_j$ are replaced by kinetic
ones $\pi_j = -i \nabla_j$, with $\nabla_j$ a connection on the torus with
non-vanishing curvature such that
\beq\label{eq:kin-mom-CR}
	[\pi_1, \pi_2] = i B.
\eeq
On $E_{n,m}$ one can always construct connections with constant curvature:
\beq
	\nabla_1^{(0)} = \frac{i m}{n + m \theta} x, \qquad \nabla_2^{(0)} =
	\frac{1}{2 \pi} \frac{\pd}{\pd x},
\eeq
satisfying $[\nabla_k^{(0)}, U_l] = i \delta_{k,l} U_l$.

Next consider an algebra automorphism $T_\alpha$ of translations by $2 \pi
\alpha_1$ and $2 \pi \alpha_2$ on $\torus$, defined by
\bals
	U_1 & \to T_\alpha (U_1) = e^{2 \pi i \alpha_1} U_1 \\
	U_2 & \to T_\alpha (U_2) = e^{2 \pi i \alpha_2} U_2.
\eals
These automorphisms are inner iff $\alpha_i = k_i \theta + n_i, \, k_i, n_i \in
\Z$. On $E_{n,m}$ this automorphism is represented by the adjoined action of the
operator $T(\alpha)$, with
\beq
	(T(\alpha) \phi)_j = e^{-2 \pi i \alpha_1 \frac{m}{n + m \theta} x} \phi_j
	(x-\alpha_2).
\eeq
The action on the connections $\nabla_i^{(0)}$ is given by
\bals
	\nabla_1^{(0)} & \to \nabla_1^{(\alpha)} = T_\alpha(\nabla_1^{(0)}) =
	\nabla_1^{(0)} + i \alpha_1 \frac{m}{n + m \theta}, \\
	\nabla_2^{(0)} & \to \nabla_2^{(\alpha)} = T_\alpha(\nabla_2^{(0)}) =
	\nabla_2^{(0)} - i \alpha_2 \frac{m}{n + m \theta}.
\eals

A review of the $\torus$-modules $E_{n,m}$ can be found in
\cite{Konechny:2000dp} for example. Using Eq. \eqref{eq:kin-mom-CR} one easily
calculates a $\theta$-deformed quantization condition for the magnetic field
strength $B$:
\beq
	2 \pi B = \frac{m}{n + m \theta}.
\eeq

\section{Representation of the Projected $U_i$}\label{sec:proj-Enm}
Using the representation of the kinetic momenta $\pi_j = -i
\nabla_j^{(\alpha)}$ on $\torus$, introduced in the previous section, we can
calculate the eigenfunctions $\psi_{\mu,j} (x)$ of the free Landau Hamiltonian
\beq\label{eq:free-Ham}
	\mathcal H_0 = \frac 1 2 ( \pi_1^2 + \pi_2^2),
\eeq
\bal
	\psi_{\mu,j} (x) & =: \langle x| \mu, j\rangle =: \psi_\mu (x) \otimes \hat
	e_j \nonumber \\
	& = \mathcal N_\mu e^{-2 \pi i \alpha_1 \frac{m}{n + m \theta} x} \phi_\mu
	\big(\sqrt{\frac{2 \pi m}{n + m \theta}} (x - \alpha_2)\big) \otimes \hat
	e_j,
\eal
with $\phi_\mu (x) = e^{-x^2/2} H_\mu (x), \mathcal N_\mu = \sqrt{2^{\mu-1}
\mu!} (2 (n/m + \theta))^{-1/4}, H_\mu (x)$ the $\mu$th Hermite polynomial and
$\hat e_j = \hat e_{j+m}$ the $(j \bmod m)$th unit vector of $\R^m$. The
eigenfunctions are orthonormal with respect to the scalar product
\beq
	\langle \mu, j| \mu', j'\rangle = \big(\int_{-\infty}^\infty dx \,
	\psi_\mu^\ast (x) \psi_{\mu'} (x) \big) \otimes (\hat e_j \cdot \hat e_{j'})
	= \delta_{\mu,\mu'} \delta_{j,j'}.
\eeq

Since $\psi_{\mu,j+m} (x) = \psi_\mu (x)$ each Landau level $\mu$ is spanned by
$m$ orthonormal eigenfunctions $\psi_{\mu,j} (x), j= 0, \dots m-1$, \ie each
wave-function $\psi_\mu (x)$ of the $\mu$th Landau level can be represented as
an $m$-dimensional vector $\vec c = (c_0, c_1, \dots c_{m-1})$ by $\psi_\mu (x)
= \sum_{j=0}^{m-1} x_j \langle x| \mu, j\rangle$.

For fixed $\alpha_i$ and using the projector $P_\mu = \sum_{j=0}^{m-1}
|\mu,j\rangle\langle\mu,j|$ onto the $\mu$th Landau level we get an
$m$-dimensional representation $\rho^{(m,n)}$ of the projected generators of
$\torus$, $U_i^{(\mu)} = P_\mu U_1 P_\mu$:
\bse\label{eq:rho}\bal
	\rho^{(m,n)} (U_1^{(\mu)})_{j, j'} & = c_\mu (\alpha_1)
		\delta_{j, (j'+1)\bmod m} \\
	\rho^{(m,n)} (U_2^{(\mu)})_{j, j'} & = c_\mu (\alpha_2)
		e^{-2 \pi i \frac n m j} \delta_{j,j'},
\eal\ese
with $c_\mu (\alpha) = e^{-1/(4 B) + 2 \pi \alpha} L_\mu (1/(2 B))$ and $L_\mu$
the $\mu$th Laguerre polynomial. The commutation relation of the projected
generators $U_i^{(\mu)}$ then reads
\beq\label{eq:CR-proj_Ui}
	U_1^{(\mu)} U_2^{(\mu)} = e^{2 \pi i \frac n m} U_2^{(\mu)} U_1^{(\mu)} =
	e^{\frac i B - 2 \pi i \theta} U_2^{(\mu)} U_1^{(\mu)}.
\eeq

The representation of the (unprojected) $U_i$ is infinite dimensional since
$\theta$ is irrational. There are infinitely many Landau levels, labeled by
$\mu$, each of which is $m$-fold degenerated. By projecting onto a finite
dimensional subspace of the representation space, \ie onto one Landau level,
the $U_i^{(\mu)}$ become finite dimensional matrices with some modified
commutation relation \eqref{eq:CR-proj_Ui}.

The representation $\rho^{(m,n)}$ is irreducible iff $m$ and $n$ are relatively
prime. For $\gcd (m,n) = d$, $m' = m/d$ and $n = n/d$, $\rho^{(m,n)}$
decomposes into $d$ $m'$-dimensional representations $\rho_0^{(m',n';j/d)}$,
with
\bse\label{eq:rho_0}\bal
	\rho_0^{(m',n';\lambda)} (U_1^{(\mu)}) & = e^{2 \pi i \frac{\lambda}{m'}}
		\rho^{(m',n')} (U_1^{(\mu)}) \\
	\rho_0^{(m',n';\lambda)} (U_2^{(\mu)}) & = \rho^{(m',n')} (U_2^{(\mu)})
\eal\ese
and $j = 0, \dots d-1$.

\section{Different Representation of $U_i$}\label{sec:rep-R2}
Next we consider a different representation of $\torus$ on the space of smooth
functions over $\R^2$ motivated by physics, with
\beq\label{eq:rep_Ui_R2}
	(U_1 \psi) (x, y) = e^{i x} \psi (x, y - \pi \theta), \qquad\text{and}\qquad
	(U_2 \psi) (x, y) = e^{i y} \psi (x + \pi \theta, y).
\eeq
We require special boundary conditions on $\psi$:
\bse\label{eq:BC_R2}\bal
	\psi (x + 2 \pi n, y) & = e^{2 \pi i \delta_1} \psi (x,y),
	\label{eq:BC1_R2}\\
	\psi (x, y + 2 \pi (n + \frac{m \theta}{2})) & = e^{2 \pi i \delta_2 - i
	m x} \psi (x,y).\label{eq:BC2_R2}
\eal\ese
Eqs. \eqref{eq:BC_R2} result from the study of magnetic translation operators
(see below).

The kinetic momenta $\pi_i$ will be represented (up to some gauge
transformation) as
\bse\label{eq:kin-mom_R2}\bal
	(\pi_1 \psi) (x,y) & = (-i (1 - \pi \theta B) \pd_x + B y) \psi (x,y)
	\qquad\text{and}\\
	(\pi_2 \psi) (x,y) & = -i \pd_y \psi (x,y).
\eal\ese

In the representation \eqref{eq:rep_Ui_R2}, the commutant of $\torus$ is
generated by four elements $Z_i, i= 1, \dots 4$ represented as
\bse\label{eq:Zi-rep_R2}\bal
	(Z_1 \psi) (x,y) & = e^{\frac i n x} \psi (x, y + \frac{\pi \theta}{n}), \\
	(Z_2 \psi) (x,y) & = e^{-\frac{i}{n + m \theta} y} \psi (x + \frac{\pi
		\theta}{n + m \theta}, y), \\
	(Z_3 \psi) (x,y) & = \psi (x + 2 \pi, y) \quad\text{and} \\
	(Z_4 \psi) (x,y) & = e^{i \frac m n x} \psi (x, y + 2 \pi (1 + \frac{m 
		\theta}{2 n})),
\eal\ese
which fulfill the commutation relations $Z_k Z_l = e^{2 \pi i \Theta_{k,l}} Z_l
Z_k$, with
\beq\label{eq:Theta}
	\Theta = \frac 1 n \begin{pmatrix}
		0 & -\frac{\theta}{n + m \theta} & -1 & 0 \\
		\frac{\theta}{n + m \theta} & 0 & 0 & 1 \\
		1 & 0 & 0 & -m \\ 0 & -1 & m & 0
	\end{pmatrix}.
\eeq
The entries of the matrix $\Theta$ have to be taken $\bmod \Z$, since they only
appear in the exponent. 

The magnetic translation operators $T_i, i=1, 2$ are chosen such that they
leave the Hamiltonian \eqref{eq:free-Ham} invariant, \ie they must commute with
$\pi_i$. In the present gauge \eqref{eq:kin-mom_R2} this gives two generators
of magnetic translations by the periods of the torus, $T_1 = Z_3$ and $T_2 =
Z_4$.

Using a potential generated by $U_i$, the Hamiltonian and thus the physical
setup is invariant under a translation by $2 \pi n_1$ and $2 \pi n_2$,
respectively. Therefore we have $T_i^{n_i} \sim \mathbb I$. This physical
requirement yields a quantization of the magnetic flux per unit cell:
\beq
	2 \pi \frac{2 \pi n B}{1 - 2 \pi B \theta} = 2 \pi m
	\qquad\Rightarrow\qquad
	2 \pi B = \frac{m}{n + m \theta},
\eeq
with some integer $m$ and $n = \gcd (n_1, n_2)$. Using this quantization
condition we see that $T_1^n$ and $T_2^n$ commute with all the other operators
and thus lie in the center of $\torus$. From the form of $\Theta$ in Eq.
\eqref{eq:Theta} we see that for $\theta$ irrational, the center of $\torus$ is
generated by these two operators. In the representation \eqref{eq:Zi-rep_R2},
$T_1^n = e^{2 \pi i \delta_1} \mathbb I$ and $T_2^n = e^{2 \pi i \delta_2}
\mathbb I$ due to the boundary conditions \eqref{eq:BC_R2}. Thus the
representation of the center of $\torus$ is trivial for $\theta$ irrational.
For $\theta \in \Q$, the center of $\torus$ is generated by $T_1^n (=Z_3^n),
T_2^n (=Z_4^n), Z_1^q$ and $Z_2^q$, where $q$ is some integer depending on
$\theta, m$ and $n$. Thus the center no longer has a trivial representation.

The energy spectrum of the free Landau-Hamiltonian \eqref{eq:free-Ham} with the
kinetic momenta $\pi_i$ of \eqref{eq:kin-mom_R2} is given by $\varepsilon_\mu =
B (\mu + \frac 1 2)$, where $\mu \in \N$ labels the Landau levels. The
corresponding eigenfunctions of the $\mu$th Landau level have to be a
superposition of functions $\psi_{\mu,k}$
\beq
	\psi_{\mu,k} (x,y) = \mathcal N_\mu e^{i k x} e^{-\frac B 2 (y + (\frac 1 B
	- \pi \theta) k)^2} H_\mu (\sqrt B (y + (\frac 1 B - \pi \theta) k)),
\eeq
with $\mathcal N_\mu = (2^{2 \mu} \pi \mu!^2/B)^{-1/4}$.

Using the first boundary condition \eqref{eq:BC1_R2}, we get $k= (l +
\delta_1)/n$, with $l \in \Z$. Thus any wave-function $\Psi (x,y)$ can be
written as
\beq
	\Psi (x,y) = \sum_{l=-\infty}^\infty c_l e^{i \frac{l+\delta_1}{n} x}
	\phi_l (y),
\eeq
with $\phi_l (y) = \phi (\sqrt B (y + (\frac 1 B - \pi \theta)
\frac{l+\delta_1}{n})) = \phi_{l+nm} (y - n \Lambda)$. The second boundary
condition \eqref{eq:BC2_R2} gives $c_{l+nm} = e^{2 \pi i \delta_2} c_l$.
Replacing $c_l = e^{-\frac{2 \pi i}{m n} l \delta_2} d_l$, the $\mu$th Landau
level is spanned by $m n$ eigenfunctions
\beq\begin{split}
	\psi_{\mu,r} (x,y) =: \langle x, y| \mu, r\rangle = \mathcal N_\mu
	\sum_{k=-\infty}^\infty & e^{-2 \pi i \frac{k m n + r}{m n} \delta_2}
	e^{i (k m + \frac{r + \delta_1}{n}) x} \\
	& \phi_\mu \big(\sqrt B (y + (\frac1 B - \pi \theta) \frac{k m n + r
	+ \delta_1}{n})\big),
\end{split}\eeq
with $r = 0, \dots mn - 1, \mathcal N_\mu = \sqrt{2^{\mu-1} \mu!}
(2(n/m+\theta))^{-1/4}, \phi_\mu (y) = e^{-y^2/2} H_\mu (y)$ and $H_\mu (y)$
the $\mu$th Hermite polynomial. These eigenfunctions are orthonormal with
respect to the scalar product
\beq
	\langle \mu, r| \mu', r'\rangle = \int_{- n \pi}^{n \pi} \frac{dx}{2 \pi n}
	\int_0^{n \Lambda} dy \, \psi_{\mu, r}^\ast (x,y) \psi_{\mu',r'} (x,y) =
	\delta_{\mu,\mu'} \delta_{r,r'}.
\eeq

Translations by $\vec a = (a_x, a_y)$ on the torus are given by an operator
$T(\vec a)$ with $T(\vec a) U_j = e^{i a_j} U_j T(\vec a)$ and commuting with
the kinetic momenta. In the presence of a magnetic field $B$, these two
conditions on commutation relations with $U_i$ and $\pi_i$ given in
\eqref{eq:rep_Ui_R2} and \eqref{eq:kin-mom_R2}, respectively, yield a
representation of the magnetic translation operator of the form
\beq
	(T(\vec a) \psi) (x,y) = e^{-i \frac{B}{1 - 2 \pi \theta B} a_y x} \psi (x
	- a_x, y - \frac{1 - \pi \theta B}{1 - 2 \pi \theta B} a_y).
\eeq

For $B \ne 0$ translations by an arbitrary vector $\vec a$ do not leave the
space of sections, satisfying \eqref{eq:BC_R2} for fixed $\delta_i$, invariant
(as this would be the case for $B = 0$), as can be seen from the commutation
relation
\beq
	T_1^{n k_1} T_2^{n k_2} T(\vec a) = e^{-i m (a_y k_1 - a_x k_2)} T(\vec a)
	T_1^{n k_1} T_2^{n k_2},
\eeq
with $k_i \in \Z$. One rather has to demand $\vec a = \frac{2 \pi}{m} (n_1,
n_2)$ with $n_i \in \Z$ (cf. \cite[Sec. 6]{Onofri:2000hk}) for the commutative
($\theta = 0$) case).

Analog to the previous section, a projection onto the $\mu$th Landau level,
using the projection operator $P_\mu = \sum_{r=0}^{mn-1} |\mu, r\rangle \langle
\mu, r|$ yields a $mn$-dimensional representation $\tilde\rho^{(m,n)}$ of the
projected generators of $\torus$, $U_i^{(\mu)} = P_\mu U_i P_\mu$:
\bse\label{eq:rhotilde}\bal
	\tilde\rho^{(m,n)} (U_1^{(\mu)})_{j,j'} & = c_\mu (\frac{\delta_2}{m})
		\delta_{j, (j+n) \bmod mn} \\
	\tilde\rho^{(m,n)} (U_2^{(\mu)})_{j,j'} & = c_\mu (-\frac{\delta_1}{m})
		e^{-\frac{2 \pi i}{m} j} \delta_{j,j'},
\eal\ese
with $c_\mu (\alpha) = e^{-1/(4 B) + 2 \pi \alpha} L_\mu (1/(2 B))$ and $L_\mu$
the $\mu$th Laguerre polynomial. Again we choose fixed phases $\delta_i$. The
representations $\tilde\rho^{(m,n)}$ are reducible and reduce to $n$
$m$-dimensional representations $\tilde\rho_1^{(m,n;j)}$, with
\bals
	\tilde \rho_1^{(m,n;j)} (U_1^{(\mu)}) & = \rho^{(m,n)} (U_1^{(\mu)}), \\
	\tilde \rho_1^{(m,n;j)} (U_2^{(\mu)}) & = e^{2\pi i \frac j m} \rho^{(m,n)}
	(U_2^{(\mu)}),
\eals
$j= 0, \dots n-1$ and $\rho^{(m,n)}$ given in \eqref{eq:rho}. To be consistent
with the definition of $\rho^{(m,n)}$ we have to replace $\alpha_1 \to
\delta_2/m$ and $ \alpha_2 \to -\delta_1/m$. For the sake of simplicity we set
$c_\mu(\alpha) = 1$, keeping in mind, that we have to reinsert this factor at
the end of the calculations.

For $m$ and $n$ relatively prime, the representations $\tilde\rho_1^{(m,n;j)}$
are irreducible and unitary equivalent to $\rho^{(m,n)}$. For $\gcd (m,n) = d$,
$m' = m/d$ and $n' = n/d$, $\tilde\rho_1^{(m,n;j)}$ decomposes into $d$
$m'$-dimensional representations $\tilde\rho_1^{(m',n';j/d,j'/d)}$, with
\bals
	\tilde\rho_0^{(m',n';\lambda,\lambda')} (U_1^{(\mu)}) & = e^{2 \pi i
		\frac{\lambda'}{m'}} \tilde\rho_1^{(m',n';\lambda)} (U_1^{(\mu)}) = 
		e^{2 \pi i \frac \lambda m} \rho_0^{(m',n';\lambda'-\lambda)}
		(U_1^{(\mu)})\\
	\tilde\rho_0^{(m,n;\lambda,\lambda')} (U_2^{(\mu)}) & = 
		\tilde\rho_1^{(m',n';\lambda)} (U_2^{(\mu)}) = 
		e^{2 \pi i \frac \lambda m} \rho_0^{(m,n;\lambda'-\lambda)}
		(U_2^{(\mu)})
\eals
and $j = 0, \dots d n -1$ and $j' = 0, d-1$. It is easy to see that
$\tilde\rho_1^{(m,n;\lambda,\lambda')} \cong \tilde\rho_1^{(m,n;\lambda + 1,
\lambda')}$. Thus any (reducible) representation $\tilde\rho^{(md, nd)}$, with
$m$ and $n$ relatively prime, decomposes into a direct sum of irreducible
representations $\rho_0^{(m,n;\lambda)}, \lambda \in \Q$:
\beq
	\tilde\rho^{(md,nd)} \cong \bigoplus_{\nu=0}^{n-1} \bigoplus_{j=0}^{d-1}
	e^{2 \pi i \frac{j}{m d}} \bigoplus_{j'=0}^{d-1} \rho_0^{(m,n;\frac{j'}{d})},
\eeq
with $\rho_0^{(m,n;\lambda)}$ given in \eqref{eq:rho_0}.

\section{Energy Corrections Due to a Periodic Potential}\label{sec:V_mu}
Using the results of the previous sections, we calculate the energy corrections
to the $\mu$th Landau level due to a small periodic perturbation $V(x_1, x_2)$
of the free Hamiltonian \eqref{eq:free-Ham}. Provided the perturbation is small
compared to the energy gap between two different Landau level, \ie $V$ does not
mix between states of two different Landau levels, we can use degenerate
perturbation theory up to first order. The corrections to the $\mu$th Landau
level are obtained by the eigenvalues of $V$ projected onto this level.

Assume a simple periodic potential
\beq\label{eq:perturbation}
	V (x_1, x_2) = 2 v (\cos x_1 + \cos x_2) = v (U_1 + U_1^\dagger + U_2 +
	U_2^\dagger ).
\eeq
Using the representation $\rho^{(m,n)}$ of \eqref{eq:rho}, the projected
potential $V^{(\mu)} = P_\mu V P_\mu$ is represented as an $m \times
m$-dimensional matrix $V^{(\mu;m,n)} := \rho^{(m,n)} (V^{(\mu)})$. For
$\gcd(m,n) = d > 1$, this matrix decomposes into $d$ $\frac m d$-dimensional
matrices $V_0^{(\mu;m',n';j/d)} := \rho_0^{(m',n';j/d)} (V^{(\mu)})$, with
$m'=m/d$ and $n'=n/d$. In the following we always assume $m$ and $n$ relatively
prime and write $m d$ and $n d$ if we want to express that they have a common
divisor $d$. The matrices $V_0^{(\mu;m',n';j/d)}$, $j=0, \dots d-1$, have the
explicit form
\begin{multline}\label{eq:V_Enm}
	(V_0^{(\mu;m,n;\lambda)})_{r,r'} = c_\mu (e^{2 \pi i (\frac \lambda m +
		\alpha_1)} \delta_{r,(r'+1) \bmod m} + e^{-2 \pi i (\frac \lambda m +
		\alpha_1)} \delta_{r,(r'-1) \bmod m} \\
		+ 2 \cos (2 \pi \frac n m r - 2 \pi \alpha_2) \delta_{r,r'}),
\end{multline}
with $r, r' = 0, \dots m-1$, $c_\mu = e^{-1/(4 B)} L_\mu (1/(2 B))$ an $L_\mu$
the $\mu$th Laguerre polynomial. According to Wilkinson \cite{Wilkinson:1984pr}
the eigenvalues $\varepsilon$ of $V_0^{(\mu;m',n';j/d)}$ are obtained by an
equation of the form
\beq\label{eq:P_Enm}
	P(\varepsilon) = \cos 2 \pi (\frac j d + m \alpha_1) + \cos 2 \pi m
	\alpha_2,
\eeq
with $P(\varepsilon)$ some $m$th order polynomial, independent of $\alpha_i$ and
$j$.

Using the representation $\tilde\rho^{(m,n)}$ of \eqref{eq:rhotilde}, the
corresponding $d^2 m n$-dimensional matrix $\tilde V^{(\mu;m d ,n d)} :=
\tilde\rho^{(md,nd)} (V^{(\mu)})$ decomposes into $d^2 n$ $m$-dimensional
matrices $\tilde V_0^{(\mu;m',n';j/d,j'/d)}$, $j, j'=0, \dots d-1$, with
\begin{multline}\label{eq:V_R2}
	(\tilde V_0^{(\mu;m,n;j/d,j'/d)})_{r,r'} = c_\mu (
		e^{\frac{2 \pi i}{m d} (j + \delta_2)} \delta_{r,(r'+1) \bmod m} +
		e^{-\frac{2 \pi i}{m d} (j + \delta_2)} \delta_{r,(r'-1) \bmod m} \\
		+ 2 \cos (2 \pi \frac n m r - 2 \pi \frac{j'-\delta_1}{m d})
		\delta_{r,r'}).
\end{multline}
Using the result of Wilkinson \cite{Wilkinson:1984pr} again, one gets an
equation of the form
\beq\label{eq:P_R2}
	P(\varepsilon) = \cos \frac{2 \pi}{d} (j + \delta_2) + \cos \frac{2 \pi}{d} 
	(j' - \delta_1).
\eeq

From Eq. \eqref{eq:P_R2} it follows that we can get rid of rational phases
$\delta_i = p_i/q_i$ in the boundary conditions \eqref{eq:BC_R2}, choosing a
$q_i$-times enlarged super-cell, \ie replacing $d \to q_i d$ and choosing $j =
p_1$ and $j' = p_2$, respectively, which is clear from a physical point of
view.

\begin{figure}[b]
	\hspace*{\fill}\includegraphics[angle=90,width=0.4\textwidth]{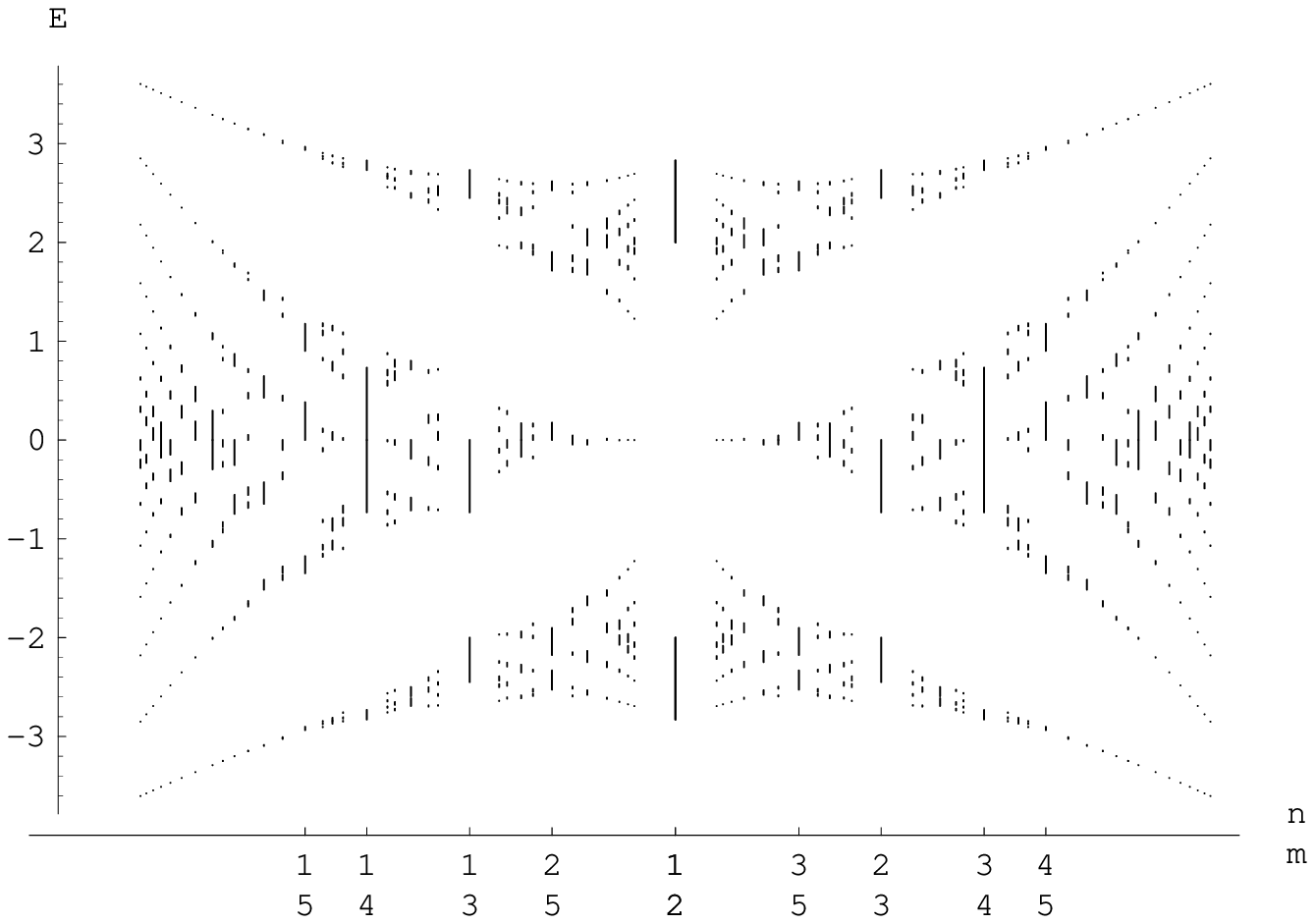}
	\hfill\hfill\includegraphics[angle=90,width=0.4\textwidth]{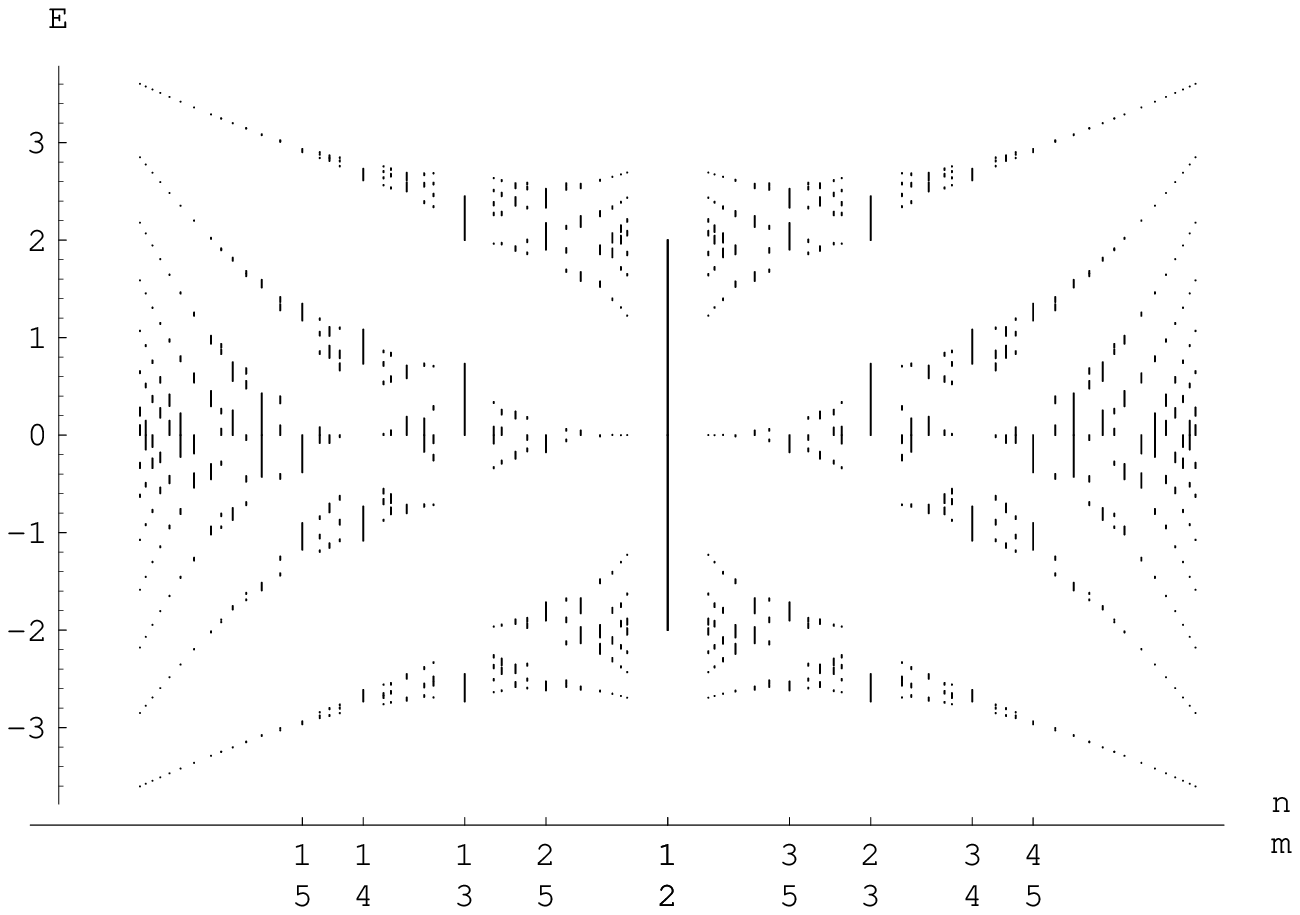}
	\hspace*{\fill}\\
	\hspace*{\fill} (a) \hfill\hfill (b) \hspace*{\fill}
	\caption{\label{fig:butterfly_part} Spectrum of $V^{(\mu;md,nd)}$ in units
	of the band width $v e^{-\frac{1}{4 B}} L_\mu (\frac{1}{2 B})$ for $m = 1,
	\dots 15, n = 1,\dots m, d \to \infty$ and $\alpha_2 = 0$ (a) and
	$\alpha_2 =\frac{1}{2 m}$ (b).}
\end{figure}

There is another fact we want to point out. To get the energy corrections to
the $\mu$th Landau level due to the perturbation \eqref{eq:perturbation} for a
fixed magnetic field, \ie a fixed ratio $\frac m n$, one has to calculate all
eigenvalues of the matrices $V^{(\mu;m d,n d)}$ and $\tilde V^{(\mu;m d,n d)}$,
respectively, for $d = 1, 2, \dots$. Using representation \eqref{eq:V_R2} and
taking $j, j' = 0, \dots d-1$, with $d \to \infty$, the (rational part of) the
Hofstadter butterfly spectrum is obtained. By the representation
\eqref{eq:V_Enm} we take only $j = 0, \dots d-1$, $d \to \infty$ and obtain
only half of the values of $P(\varepsilon)$ of Eq. \eqref{eq:P_R2}. This is
shown in Fig. \ref{fig:butterfly_part} for $\alpha_2 = 0$ (a) and $\alpha_2 =
\frac{1}{2 m}$ (b). Since we imposed toroidal boundary conditions on the
configuration space, the magnetic flux is some rational number. So only the
rational part, \ie the part belonging to rational fluxes of the butterfly
arises.

\begin{figure}[b]
	\hspace*{\fill}\includegraphics[width=0.9\textwidth]{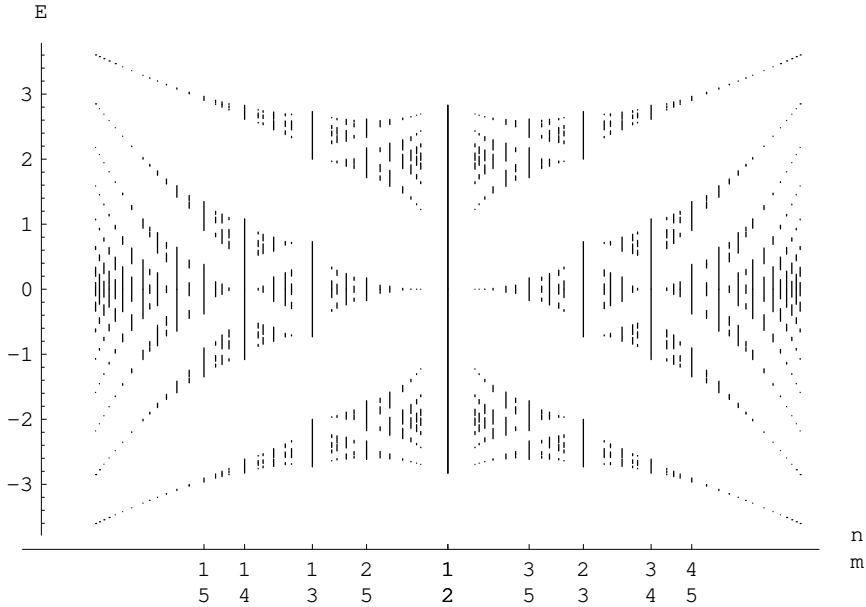}
	\hspace*{\fill}
	\caption{\label{fig:butterfly} Spectrum of $\tilde V^{(\mu;md,nd)}$
	in units of the band width $v e^{-\frac{1}{4 B}} L_\mu (\frac{1}{2 B})$ for
	$m=1,\dots 15, n = 1,\dots m$ and $d \to \infty$.}
\end{figure}

In either representation, the spectrum for fixed $\alpha_i$, $\delta_i$ is a
pure point spectrum, which splits into $m$ separate parts, in each of which the
eigenvalues $\varepsilon$ lie dense (for $d \to \infty$). In Fig.
\ref{fig:butterfly} the spectrum of $\tilde V^{(\mu;m d,n d)}$ for $d \to
\infty$ is plotted in units of the band width, $v e^{-\frac{1}{4 B}} L_\mu
(\frac{1}{2 B})$. One sees that the $\theta$-deformation of the underlying
manifold has no effect on the spectrum (besides a rescaling $B \to \frac{B}{1 -
2 \pi \theta B}$).

\end{document}